\documentclass[
,final            
,numberedheadings 
]
{aipproc}
\layoutstyle{6x9}
\newcommand{\sr}{$\cal S$}
\newcommand{\ps}{$\cal P$}
\newcommand{\ve}{$\cal V$}
\newcommand{\ax}{$\cal A$}
\newcommand{\oc}{$\cal O$}
\newcommand{\de}{$\cal D$}
\newcommand{\ds}{\displaystyle}
\newcommand{\lsm}{L$\sigma$M}

\newcommand{\refc}[1]{Ref.~\cite{#1}}
\newcommand{\ha}{\frac{1}{2}}

\newcommand{\be}{\begin{equation}}
\newcommand{\ee}{\end{equation}}
\newcommand{\bea}{\begin{eqnarray}}
\newcommand{\eea}{\end{eqnarray}}
\newcommand{\eqr}[1]{Eq.~(\ref{#1})}
\newcommand{\eqrs}[2]{Eqs.~(\ref{#1}) and (\ref{#2})}

\begin{document}
\title{$SU(3)$ Mass Splittings for $\bar{q}q$ Mesons and  $qqq$ Baryons}
\author{M.\ D.\ Scadron}{
address={Physics Department, University of Arizona, Tucson, AZ 85721, USA}
}
\author{G.\ Rupp}{
address={Centro de F\'{\i}sica das Interac\c{c}\~{o}es Fundamentais,
Instituto Superior T\'{e}cnico, Lisbon, Portugal}
}
\author{E.\ van Beveren}{
address={Departamento de F\'{\i}sica, Universidade de Coimbra,
Coimbra, Portugal}
}
\author{F.\ Kleefeld}{
address={Centro de F\'{\i}sica das Interac\c{c}\~{o}es Fundamentais,
Instituto Superior T\'{e}cnico, Lisbon, Portugal}
}

\begin{abstract}
By comparing $SU(3)$-breaking scales of linear mass formulae, it is shown that
the lowest vector, axial-vector, and scalar mesons all have a $\bar{q}q$
configuration, while the ground-state octet and decuplet baryons are $qqq$.
Also, the quark-level linear $\sigma$ model is employed to predict similar
$\bar{q}q$ and $qqq$ states. Finally, the approximate mass degeneracy of the
scalar $a_0$(980) and $f_0$(980) mesons is demonstrated to be accidental.
\end{abstract}

\maketitle

\section{Introduction}
In the quark model, one usually assumes that pseudoscalar (\ps),
vector (\ve), and axial-vector (\ax) mesons are $\bar{q}q$, whereas
octet (\oc) and decuplet (\de) baryons are $qqq$ states. However,
it is often argued \cite{PDG02} that the light scalar (\sr) mesons are
non-$\bar{q}q$ candidates, in view of their low masses. In this short paper, we
shall show that the ground-state meson nonets \ps, \sr, \ve, and \ax\ are \em
all \em \/$\bar{q}q$, hence including the light scalars, while the lowest \oc\
and \de\ baryons are $qqq$ states. 

In Sec.~2, $SU(3)$ mass splittings for loosely bound \ve, \ax, and \sr\ states
are shown to have symmetry-breaking scales of 16\%, 8\%, and 25\%,
respectively, using linear mass formulae. We apply the latter formulae to
$qqq$ \oc\ and \de\ states in Sec.~3, leading to $SU(3)$-breaking scales of
13\% and 12\%, respectively. Then in Sec.~4, we employ the quark-level linear
$\sigma$ model (\lsm) to predict similar $\bar{q}q$ and $qqq$ states as in
Secs.~2 and 3. Next in Sec.~5, we study the \sr\ $\bar{q}q$ states and argue
why the \ve\ states have slightly higher masses, on the basis of the
nonrelativistic quark model. Moreover, the approximate mass degenaracy of the
\sr\ $a_0$(980) and $f_0$(980) mesons is shown to be just accidental. We
summarize our results in Sec.~6.

\section{Mass splittings for $SU(3)$ \ve, \sr, \ax\ ground states}

Although meson masses are expected to appear \em quadratically \em \/in model
Lagrangians, while they must appear so for \ps\ states \cite{PS80}, for \ve,
\sr, \ax\ states a Taylor-series linear form for $SU(3)$ mass splittings is
also possible. Thus consider a Hamiltonian density $H=H(\lambda_0)+
H_{ss}(\lambda_8)$ using Gell-Mann matrices. Then the vector-meson-nonet masses
$m_{\cal V}=m_{\cal V}^0-\delta m_{\cal V}\,d_{\bar{i}8i}$ are
\be
\begin{array}{ccccr} 
m_{\rho,\omega}&=&m_{\cal V}^0\,-\,\displaystyle\frac{\delta m_{\cal V}}
{\sqrt{3}}&\approx&776\;\mbox{MeV} \; ,\\[4mm]
m_{K^*}&=&m_{\cal V}^0\,+\,\displaystyle\frac{\delta m_{\cal V}}{2\sqrt{3}}&
\approx&892\;\mbox{MeV} \; , \\[4mm]
m_{\phi}&=&m_{\cal V}^0\,+\,\displaystyle\frac{2\delta m_{\cal V}}{\sqrt{3}}&
\approx&1020\;\mbox{MeV} \; ,
\end{array}
\label{vector}
\ee
with $\phi\approx\bar{s}s$. Measured vector masses \cite{PDG02} suggest average
mass splittings
\be
m_{\cal V}^0 \; \approx \; 850\;\mbox{MeV} \;\;\; , \;\;\; \delta m_{\cal V} \;
\approx \; 140\;\mbox{MeV} \;\;\; ,
\label{mv}
\ee
giving an $SU(3)$-breaking scale of
$\delta m_{\cal V}/m_{\cal V}^0\approx16\%$.

Likewise, the ground-state axial-vector mesons (while slightly ambiguous) still
suggest
\be
\begin{array}{ccccl}
m_{a_1(1260),\:f_1(1285)}&=&m_{\!\cal A}^0\,-\,\displaystyle
\frac{\delta m_{\!\cal A}}{\sqrt{3}}& \approx&1256\;\mbox{MeV} \; , \\[4mm] 
m_{K_1(1270)}&=&m_{\!\cal A}^0\,+\,\displaystyle
\frac{\delta m_{\!\cal A}}{2\sqrt{3}}& \approx&1273\;\mbox{MeV} \; , \\[4mm] 
m_{f_1(1420)}&=&m_{\!\cal A}^0\,+\,\displaystyle\frac{2\delta m_{\!\cal A}}
{\sqrt{3}}&\approx&1426\;\mbox{MeV} \; .
\end{array}
\label{axial}
\ee
Here, we assume the $f_1$(1420) is mostly $\bar{s}s$, because the PDG
\cite{PDG02} reports $f_1(1420)\to KK\pi, \,K^*K$ as dominant, while $f_1(1285)
\to KK\pi, \,K^*K$ are almost absent. Thus, $f_1(1285)$ is mostly $\bar{n}n$,
like the nonstrange $a_1(1260)$ (with $a_1\to\sigma\pi$ seen, but $a_1\to
f_0(980)\pi$ not seen, because $f_0(980)$ is mostly $\bar{s}s$). Then the
pattern of Eqs.~(\ref{axial}) suggests approximate average mass splittings
\be
m_{\!\cal A}^0 \; \approx \; 1305\;\mbox{MeV} \;\;\;,\;\;\; \delta m_{\!\cal A}
\; \approx \; 98\;\mbox{MeV} \;\;\;,\;\;\;
\frac{\delta m_{\!\cal A}}{m^0_{\!\cal A}} \; \approx \; 8\% \;\;\;.
\label{ma}
\ee

Also the scalar masses (not incompatible with \refc{PDG02}) predicted from
the \lsm\ discussed in Sec.~4  obey the mass-splitting pattern
\be
\begin{array}{ccccl} 
m_{\sigma_n}&=&m_{\!\cal S}^0\,-\,\displaystyle\frac{\delta m_{\!\cal S}}
{\sqrt{3}}&\approx&650\;\mbox{MeV} \; , \\[4mm]
m_{\kappa}&=&m_{\!\cal S}^0\,+\,\displaystyle\frac{\delta m_{\!\cal S}}
{2\sqrt{3}}&\approx&800\;\mbox{MeV} \; , \\[4mm]
m_{\sigma_s}&=&m_{\!\cal S}^0\,+\,\displaystyle\frac{2\delta m_{\!\cal S}}
{\sqrt{3}}&\approx&970\;\mbox{MeV} \; .
\end{array}
\label{scalar}
\ee
Here, $m_{\sigma_n(650)}$ is near the PDG average \cite{PDG02}
$m_{f_0(600)}$, \,$m_{\kappa(800)}$ is near the E791 value \cite{A02}
$797\pm19$ MeV, and $m_{\sigma_s(970)}$ from Sec.~5 and the
Appendix is near the PDG value $m_{f_0(980)}$, which is thus mostly $\bar{s}s$.
The masses from Eqs.~(\ref{scalar}) then give the average mass splittings
\be
m_{\!\cal S}^0 \; \approx \; 753\;\mbox{MeV} \;\;\;,\;\;\; \delta m_{\!\cal S}
\; \approx \; 185\;\mbox{MeV} \;\;\;,\;\;\;
\frac{\delta m_{\!\cal S}}{m^0_{\!\cal S}} \; \approx \; 25\% \;\;\;.
\label{ms}
\ee

The fact that the $\bar{q}q$ scalars have an $SU(3)$-breaking scale of 25\%,
about double the scale of \ve\ and \ax\ ground states, further suggests that,
whereas the \ve, \ax\ are $\bar{q}q$ loosely bound states, the $\bar{q}q$ \sr\
states (with quarks touching in the NJL scheme \cite{NJL61}) are ``barely''
elementary-particle partners of the tightly bound \ps\ states (discussed in
Sec.~4).

The mean of the slightly varying \ve, \ax, \sr\ mass scales in
Eqs.~(\ref{mv},\ref{ma},\ref{ms}) is $m^0=969$ MeV, $\delta m=141$ MeV, and
the latter are close to the baryon mass-splitting scales which we derive next.

\section{Loosely bound $qqq$ baryons}

In this same Taylor-series spirit, the \oc\ baryon $SU(3)$ mass splitting 
$m_{\!\cal O}=m^0_{\!\cal O}-\delta m_{\!\cal O}(d_{ss}d^{\bar{i}8i}+
f_{ss}if^{\bar{i}8i})$ for $d_{ss}+f_{ss}=1$,
predicts (the index $ss$ means semistrong)  
\be
\begin{array}{cclcr} 
m_{N}&=&m_{\!\cal O}^0\,-\,\displaystyle\frac{\delta m_{\!\cal O}}{2\sqrt{3}}\,
(-d_{ss}+3f_{ss})&\approx&939\;\mbox{MeV} \; , \\[4mm]
m_{\Lambda}&=&m_{\!\cal O}^0\,+\,\displaystyle\frac{\delta m_{\!\cal O}}
{\sqrt{3}}\,d_{ss}&\approx&1116\;\mbox{MeV} \; , \\[4mm]
m_{\Sigma}&=&m_{\!\cal O}^0\,-\,\displaystyle\frac{\delta m_{\!\cal O}}
{\sqrt{3}}\,d_{ss}&\approx&1193\;\mbox{MeV} \; , \\[4mm]
m_{\Xi}&=&m_{\!\cal O}^0\,+\,\displaystyle\frac{\delta m_{\!\cal O}}{2\sqrt{3}}
\,(d_{ss}+3f_{ss})&\approx&1318\;\mbox{MeV} \; . 
\end{array}
\label{octet}
\ee
The $(d/f)_{ss}$ ratio can be found from Eqs.~(\ref{octet}) as
\be
\left(\frac{d}{f}\right)_{\!\!\!ss}\;=\;-\frac{3}{2}\:\frac{m_\Sigma-m_\Lambda}
{m_\Xi-m_N} \; \approx \; -0.305 \;\;\;,\;\;\; d_{ss} \; \approx \; -0.44
\;\;\;,\;\;\; f_{ss} \; \approx \; 1.44 \;\;\; .
\label{dfo}
\ee
Thus, Eqs.~(\ref{octet}) predict the average mass splittings
\be
m_{\!\cal O}^0 \; \approx \; 1151\;\mbox{MeV} \;\;\;,\;\;\; \delta m_{\!\cal O}
\; \approx \; 150\;\mbox{MeV} \;\;\;,\;\;\;
\frac{\delta m_{\!\cal O}}{m^0_{\!\cal O}} \; \approx \; 13\% \;\;\;.
\label{mo}
\ee

The $SU(3)$ \de\ baryon masses
$m_{\!\cal D}=m^0_{\!\cal D}+\delta m_{\!\cal D}$ have $m^0_{\cal D}$ weighted
by wave functions
\be
\overline{\Psi}^{(abc)}\Psi_{(abc)} \; = \; \overline{\Delta}\Delta \, + \,
\overline{\Sigma}^*\Sigma^* \, + \, \overline{\Xi}^*\Xi^* \, + \,
\overline{\Omega}\Omega \; ,
\label{dwf}
\ee
and $\delta m_{\!\cal D}$ is weighted by
\be
3\,\overline{\Psi}^{(ab3)}\Psi_{(ab3)} \; = \; \overline{\Sigma}^*\Sigma^* \, +
\, 2\,\overline{\Xi}^*\Xi^* \, + \, 3\,\overline{\Omega}\Omega \; .
\label{dmd}
\ee
Then the $SU(3)$ \de\ masses are predicted (in MeV) to be
\be
\begin{array}{cclcl}
m_\Delta & = & m^0_{\!\cal D} & \approx & 1232 \; , \\[2mm]
m_{\Sigma^*} & = & m^0_{\!\cal D}\,+\,\delta m_{\!\cal D} & \approx & 1385 \; ,
\;\;\; \mbox{with} \;\; \delta m_{\!\cal D} \; \approx \; 153 \; , \\[2mm]
m_{\Xi^*} & = & m^0_{\!\cal D}\,+\,2\delta m_{\!\cal D} & \approx & 1533 \; ,
\;\;\; \mbox{with} \;\; \delta m_{\!\cal D} \; \approx \; 151 \; , \\[2mm]
m_{\Omega} & = & m^0_{\!\cal D}\,+\,3\delta m_{\!\cal D} & \approx & 1672 \; ,
\;\;\; \mbox{with} \;\; \delta m_{\!\cal D} \; \approx \; 147 \; .
\end{array}
\label{decuplet}
\ee
This corresponds to average mass splittings
\be
m_{\!\cal D}^0 \; \approx \; 1232\;\mbox{MeV} \;\;\;,\;\;\; \delta m_{\!\cal D}
\; \approx \; 150\;\mbox{MeV} \;\;\;,\;\;\;
\frac{\delta m_{\!\cal D}}{m^0_{\!\cal D}} \; \approx \; 12\% \;\;\;.
\label{md}
\ee

It is interesting that both loosely bound $qqq$ \oc\ and \de\ symmetry-breaking
scales of about 150 MeV are near the $\bar{q}q$ \ve, \ax, \sr\ mean
mass-splitting scale of $\delta m=141$ MeV. However, the $SU(3)$-breaking scale
of 25\% for scalars
is almost double the 12--16\% scales of \ve, \ax, \oc, \de\ states. This
suggests that \ve, \ax, \oc, \de\ $\bar{q}q$ or $qqq$ states are all loosely
bound, in contrast with the $\bar{q}q$ \sr\ and, of course, the \ps\ states
(see above). In fact, the latter Nambu--Goldstone \ps\ states are massless in
the chiral limit (CL) $p^2=m^2_\pi=0$, $p^2=m^2_K=0$, as the tightly-bound
measured \cite{PDG02} $\pi^+$ and $K^+$ charge radii indicate \cite{SKRB03}.

\section{Constituent quarks and the quark-level \lsm}
Formulating the \ps\ and \sr\ $\bar{q}q$ states as elementary chiral partners
\cite{BKRS02}, the Lagrangian density of the $SU(2)$ quark-level linear
$\sigma$ model (\lsm) has, after the spontaneous-symmetry-breaking shift, the
interacting part \cite{GML60}
\begin{equation}
{\cal L}^{\mbox{\footnotesize int}}_{\mbox{\footnotesize\lsm}} = g\,\bar{\psi}
(\sigma+i\gamma_5\vec{\tau}\cdot\vec{\pi})\psi\,+\,g'\,\sigma\,(\sigma^2+\pi^2)
\,-\,\frac{\lambda}{4}\,(\sigma^2+\pi^2)^2 \, - \, f_\pi g\,\bar{\psi}\psi \; ,
\label{lsm}
\end{equation}
with tree-order CL couplings related as (for $f_\pi\approx93$ MeV)
\be
g \; = \; \frac{m_q}{f_\pi} \;\;\;\; , \;\;\;\; g' \; = \;
\frac{m^2_\sigma}{2f_\pi} \; = \; \lambda\,f_\pi \; .
\label{ggprime}
\ee
The $SU(2)$ and $SU(3)$ chiral Goldberger--Treiman relations (GTRs) are
\begin{equation}
f_{\pi}\,g \; = \; \hat{m} \;= \; \ha\,(m_u+m_d) \;\;\;\; , \;\;\;\; f_K\,g \;
= \; \ha\,(m_s+\hat{m}) \; .
\label{gtrs}
\end{equation}
Since $f_K/f_\pi\approx1.22$ \cite{PDG02}, the constituent-quark-mass ratio
from \eqr{gtrs} becomes
\be
1.22 \; \approx \; \frac{f_K}{f_\pi} \; = \; \ha\,(1+\frac{m_s}{\hat{m}})
\;\;\;\; \Rightarrow \;\;\;\; \frac{m_s}{\hat{m}} \; \approx \; 1.44 \; ,
\label{rmsmh}
\ee
which is independent of the value of $g$. In loop order, Eqs.~(\ref{ggprime})
are recovered, along with \cite{DS95,SKRB03}
\be
m_\sigma \; = \; 2m_q \;\;\;\; , \;\;\;\; g \; = \; \frac{2\pi}{\sqrt{N_c}}
\;\;\;\; , \;\;\;\; \mbox{for} \;\; N_c \; = \; 3 \; .
\label{loop}
\ee
Here, the first equation is the NJL relation \cite{NJL61}, now true for the
\lsm\ as well. The second equation in \eqr{loop} was first found via the
$Z=0$ compositeness relation \cite{SWS62_98}, separating the elementary $\pi$
and $\sigma$ particles from the bound states $\rho$, $\omega$, and $a_1$.

We first estimate the (non-chiral-limiting) nonstrange and strange constituent
quark masses from the GTRs (\ref{gtrs}), together with the \lsm\ loop-order
result (\ref{loop}):
\be
\begin{array}{ccccccc}
\hat{m} & \approx & g\,f_\pi & \approx & \displaystyle\frac{2\pi}{\sqrt{3}}\:
(93\:\mbox{MeV}) & \approx & 337\;\mbox{MeV} \; , \\[2mm]
m_s & = & \left(\displaystyle\frac{m_s}{\hat{m}}\right)\hat{m} & \approx
& 1.44\,\hat{m} & \approx & 485\;\mbox{MeV} \; .
\end{array}
\ee
These quark-mass scales in turn confirm the mass-splitting scales found in
Secs.~2, 3:
\be
\delta m_{\cal V}\,=\,\delta m_{\!\cal A}\,=\,\delta m_{\!\cal O}\,=\,
\delta m_{\!\cal D} \; \approx \; (485-337)\;\mbox{MeV}\,=\,148\;\mbox{MeV}\; ,
\label{dmlsm}
\ee
near 140, 98, 150, and 150 MeV, respectively. Also the $SU(3)$ non-vanishing
masses are predicted as
\be
\begin{array}{ccccccr}
m^0_{\cal V}&=&m^0_{\!\cal A}&\approx&m_s+ \hat{m}&\approx& 822\;\mbox{MeV}\; , 
\\[2mm]
m^0_{\cal O}&=&m^0_{\!\cal D}&\approx&m_s+2\hat{m}&\approx&1160\;\mbox{MeV}\; ,
\label{mvaod}
\end{array}
\ee
near the 850, 1151, and 1232 MeV $m^0$ masses in Secs.~2, 3.

\section{\sr\ scalars and accidental degeneracies}

An almost degenerate case in the nonrelativistic quark model (NRQM) is
\cite{MSW88}, in the context of QCD,
\be
m_{\!\cal S} \; \approx \; m_{\cal V} \, - \,
\frac{2\alpha_s^{\mbox{\footnotesize eff}}}{m^2_{\mbox{\footnotesize dyn}}}\,
\left(\frac{\vec{L}\cdot\vec{S}}{r^3}\right) \;
= \; 780\;\mbox{MeV}\,-\,140\;\mbox{MeV} \; = \; 640\;\mbox{MeV} \; ,
\label{nrqm}
\ee
where the ground-state vector mesons have $L=0$ and so no spin-orbit
contribution to the mass.
This corresponds to $m_{\sigma(650)}\approx m_{\omega(782)}-140$ MeV $=642$
MeV. Equivalently, invoking the $I\!=\!1/2$ CGC of $1/2$, one predicts via the
NRQM $\,m_{\kappa(800)}\approx m_{K^*(892)}-70$ MeV $=822$~MeV. Or invoking
instead the $\bar{s}s$ CGC of $1/3$, one gets $m_{\sigma_s(970)}\approx
m_{\phi(1020)}-47$~MeV $= 973$ MeV.

However, for the elementary-particle \ps\ and \sr\ states, one should invoke
the infinite-momentum-frame (IMF, see Appendix) scalar-pseudoscalar $SU(3)$
equal-splitting laws (ESLs), reading \cite{S82_92}
\be
m^2_\sigma\,-\,m^2_\pi \; \approx \; m^2_\kappa\,-\,m^2_K \; \approx \;
m^2_{a_0}\,-\,m^2_{\eta_{\mbox{\footnotesize avg}}} \; \approx \; 0.40 \;
\mbox{GeV}^2 \; ,
\label{esls}
\ee
where $m_{\eta_{\mbox{\footnotesize avg}}}$ is the average $\eta$,
$\eta^\prime$ mass 753 MeV. These ESLs hold for $m_{\sigma(650)}=2\hat{m}$
and $m_{\kappa(800)}=2\sqrt{m_s\hat{m}}=809$ MeV, the NJL-\lsm\ values. Using
the ESLs~(\ref{esls}) to predict the $a_0$ mass, one finds
\be
m_{a_0}\;=\;\sqrt{0.40\;\mbox{GeV}^2\,+\,m^2_{\eta_{\mbox{\footnotesize avg}}}}
\; \approx \; 983.4\;\mbox{MeV} \; ,
\label{anot}
\ee
very close to the PDG value $984.7\pm1.2$ MeV. Thus, the nearness of the 
$a_0$(980) and $f_0$(980) masses, the latter scalar being mostly $\bar{s}s$
and so near the vector $\bar{s}s$ $\phi$(1020) (see above), is indeed an
accidental degeneracy. Note that a similar (approximate) degeneracy is found
in the dynamical unitarized quark-meson model of \refc{BRMDRR86}, where the
same $\bar{q}q$ assignments are employed as here.

\section{Summary and conclusions}
The usual field-theory picture is that meson masses should appear quadratically
and baryon masses linearly in Lagrangian models based on the Klein--Gordon
and Dirac equations. However, in Secs.~2 and 3 we have studied both mesons and
baryons in a {\it linear}-mass $SU(3)$-symmetry Taylor-series sense. Instead,
in Sec.~5 we have studied symmetry breaking in the IMF, with
$E=[p^2+m^2]^{1/2}\approx p\,[1+m^2/2p^2+\ldots]$. Here, between brackets, the
$1$ indicates the symmetry limit, and the \em quadratic \em \/mass term means
that both meson and baryon masses are \em squared \em \/in the mass-breaking
IMF for $\Delta S\!=\!1$ ESLs. While the former mass-splitting approach (with
linear masses) fits all \ve, \ax, \sr, \oc, and \de\ ground-state
$SU(3)$-flavor multiplets, so does the latter (with quadratic masses) for the
IMF-ESLs. Nevertheless, Nambu--Goldstone pseudoscalars \ps\ \em always \em
\/involve \em quadratic \em \/masses. Both approaches suggest that all 
ground-state mesons (\ps, \sr, \ve, \ax) are $\bar{q}q$ states, while baryons
(\oc, \de) are $qqq$ states. This picture is manifest in the quark-level \lsm\
of Sec.~4. Finally, the accidental scalar degeneracy between the $\bar{s}s$
$f_0$(980) and the $\bar{n}n$ $a_0$(980) was explained in Sec.~5, via the IMF
quadratic-mass ESLs --- also compatible with mesons being $\bar{q}q$ and
baryons $qqq$ states.

\begin{theacknowledgments}
One of the authors (M.D.S.) wishes to thank V.~Elias for useful discussions.
This work was partly supported by the
{\it Funda\c{c}\~{a}o para a Ci\^{e}ncia e a Tecnologia}
of the {\it Minist\'{e}rio da Ci\^{e}ncia e da Tecnologia} \/of Portugal,
under contract no.\ POCTI/\-FNU/\-49555/\-2002, and grant no.\
SFRH/\-BPD/\-9480/\-2002.
\end{theacknowledgments}

\appendix

\section{Kinematic infinite-momentum frame}

The infinite-momentum frame (IMF) has two virtues: \,(i) 
$\,E=[p^2+m^2]^{1/2}\approx p+m^2/2p+\ldots$, for $p\to\infty$, requires \em
squared \em \/masses when the lead term $p$ is eliminated, using $SU(3)$
formulae with coefficients $1\!+\!3=2\!+\!2$, as e.g.\ the Gell-Mann--Okubo
linear mass formula $\Sigma\!+\!3\Lambda=2N\!+\!2\Xi$, valid to 3\%; \,(ii)
when $p\to\infty$, dynamical tadpole graphs are suppressed \cite{FF65}. In
fact, $\Sigma^2\!+\!3\Lambda^2=2N^2\!+\!2\Xi^2$ is also valid empirically to
3\%.  This squared $qqq$ baryon mass formula can be interpreted as a
$\Delta S\!=\!1$ ESL, which holds for both \oc\ and \de\ baryons \cite{S82_92}:
\be
\begin{array}{ccccccccc}
\Sigma\Lambda-N^2 &\!\!\approx\!\!& \Xi^2-\Sigma\Lambda &\!\!\approx\!\! &
\ds\ha\left(\Xi^2-N^2\right) & & &\!\!\approx\; 0.43\;\mbox{GeV}^2 \; , \\[2mm]
{\Sigma^*}^2-\Delta^2 &\!\!\approx\!\!& {\Xi^*}^2-{\Sigma^*}^2 &\!\!\approx\!\!
&\Omega^2-{\Xi^*}^2&\!\!\approx\!\!& \ds\ha\,\left(\Omega^2-{\Sigma^*}^2\right)
&\!\!\approx \;0.43\;\mbox{GeV}^2 \; .
\end{array}
\label{baresls}
\ee

However, the $\bar{q}q$ pseudoscalar and vector $\Delta S\!=\!1$ ESLs have
about one half this scale (also empirically valid to 3\%), viz.\
\be
m^2_K\,-\,m^2_\pi \; \approx \; m^2_{K^*}\,-\,m^2_\rho \; \approx \;
m^2_\phi\,-\,m^2_{K^*}\; \approx \; \ha\,(m^2_\phi\,-\,m^2_\rho) \; \approx \;
0.22\;\mbox{GeV}^2 \; ,
\label{pseudoscalars}
\ee
as roughly do the $\bar{q}q$ scalars found in Sec.~2, i.e.,
\be
m^2_{\kappa(800)}\,-\,m^2_{\sigma_{n}(650)} \; \approx \; m^2_{\sigma_s(970)}
\,-\,m^2_{\kappa(800)} \; \approx \; \mbox{0.22\,--\,0.30 GeV}^2 \; .
\label{scalars}
\ee
This approximate factor of 2 between Eqs.~(\ref{baresls}) and
Eqs.~(\ref{pseudoscalars},\ref{scalars}) is because there are two
$\Delta S\!=\!1$ $qqq$ transitions, whereas there is only one $\Delta S\!=\!1$
transition for $\bar{q}q$ configurations.

So if we take \eqr{scalars} as physically meaningful, we may write
\be
2m^2_{\kappa} \; \approx \; m^2_{\sigma(600)} \,+\,m^2_{f_0(980)} \; \approx \;
m^2_{\sigma_{n}(650)} \,+\,m^2_{\sigma_{s}(970)} \; \approx \;
\mbox{1.32\,--\,1.36 GeV}^2 \; ,
\label{kappa}
\ee
yielding $m_\kappa\approx819$ MeV close to experiment, which again suggests
these scalars are $\bar{q}q$ states.

These IMF quadratic mass schemes, along with the NJL-\lsm\ $\kappa$ mass
$m_{\kappa(800)}=2\sqrt{m_s\hat{m}}=809$ MeV, again suggest (as do the
empirical scales of \eqrs{pseudoscalars}{scalars} vs. Eqs.~(\ref{baresls}))
that \em all \em \/ground-state meson nonets are $\bar{q}q$, whereas the baryon
octet and decuplet are $qqq$ states.

\end{document}